\documentclass[%
 aip,
 amsmath,amssymb,
 reprint,%
]{revtex4-1}

\usepackage{color}
\usepackage{xcolor}
\usepackage{subcaption}
\usepackage{graphicx}% Include figure files
\usepackage{dcolumn}% Align table columns on decimal point
\usepackage{bm}% bold math
\usepackage[utf8]{inputenc}
\usepackage[T1]{fontenc}
\usepackage{mathptmx}
\usepackage{etoolbox}
\usepackage{multirow}

\makeatletter
\def\@email#1#2{%
 \endgroup
 \patchcmd{\titleblock@produce}
  {\frontmatter@RRAPformat}
  {\frontmatter@RRAPformat{\produce@RRAP{*#1\href{mailto:#2}{#2}}}\frontmatter@RRAPformat}
  {}{}
}%
\makeatother
\begin{document}

\preprint{AIP/123-QED}

\title{Design and Performance of Parallel-channel Nanocryotrons in Magnetic Fields}

\author{Timothy Draher}
 \affiliation{Argonne National Laboratory, Materials Science Division, Lemont, Illinois, 60439, USA}
 \affiliation{Northern Illinois University, Department of Physics, Dekalb, Illinois, 60115, USA}
\author{Tomas Polakovic}%
 \affiliation{Argonne National Laboratory, Physics Division, Lemont, Illinois, 60439, USA}
\author{Yi Li}
 \affiliation{Argonne National Laboratory, Materials Science Division, Lemont, Illinois, 60439, USA}
\author{John Pearson}
 \affiliation{Argonne National Laboratory, Materials Science Division, Lemont, Illinois, 60439, USA}
 \affiliation{Center for Nanoscale Materials, Argonne National Laboratory, Lemont, IL 60439, USA}
\author{Alan Dibos}
 \affiliation{Argonne National Laboratory, Materials Science Division, Lemont, Illinois, 60439, USA}
 \affiliation{Center for Nanoscale Materials, Argonne National Laboratory, Lemont, IL 60439, USA}
 \author{Zein-Eddine Meziani}
 \affiliation{Argonne National Laboratory, Physics Division, Lemont, Illinois, 60439, USA}
\author{Zhili Xiao}
 \affiliation{Argonne National Laboratory, Materials Science Division, Lemont, Illinois, 60439, USA}
 \affiliation{Northern Illinois University, Department of Physics, Dekalb, Illinois, 60115, USA}
\author{Valentine Novosad}
 \affiliation{Argonne National Laboratory, Materials Science Division, Lemont, Illinois, 60439, USA}
 \affiliation{Argonne National Laboratory, Physics Division, Lemont, Illinois, 60439, USA}
\email{Author to whom correspondence should be addressed: novosad@anl.gov}

\date{\today}% It is always \today, today,
             %  but any date may be explicitly specified

\begin{abstract}

We introduce a design modification to conventional geometry of the cryogenic three-terminal switch, the nanocryotron (nTron). The conventional geometry of nTrons is modified by including parallel current-carrying channels, an approach aimed at enhancing the device's performance in magnetic field environments. The common challenge in nTron technology is to maintain efficient operation under varying magnetic field conditions. Here we show that the adaptation of parallel channel configurations leads to an enhanced gate signal sensitivity, an increase in operational gain, and a reduction in the impact of superconducting vortices on nTron operation within magnetic fields up to 1 Tesla. Contrary to traditional designs that are constrained by their effective channel width, the parallel nanowire channels permits larger nTron cross sections, further bolstering the device's magnetic field resilience while improving electro-thermal recovery times due to reduced local inductance. This advancement in nTron design not only augments its functionality in magnetic fields but also broadens its applicability in technological environments, offering a simple design alternative to existing nTron devices.
\end{abstract}

\maketitle
The superconducting nanowire cryotron (nTron) is a current-based three terminal device that is the natural evolution of the original macroscopic cryotron superconducting switch~\cite{buck1956cryotron}. Compared to the related Josephson junction-based devices, the nTron has an ultra-compact size, insensitivity to magnetic noise, and the ability to be integrated into larger superconducting and semiconductor hybrid circuits \cite{McCaughan2014, Tanaka2017, Zhao2017, Zhao2018, McCaughan2019, Zheng2020}. In addition, nTron fabrication and design can be tuned to fit a range of applications such as signal amplification and digital logic \cite{Krylov2019, Huang2023}. This makes the nTron a desirable candidate for on chip integration with superconducting nanowire detectors. As there has been a recent uptick of demand of superconducting nanowire detectors outside of the field of nanophotonics~\cite{polakovic2020unconventional,khalek2022science}, pushing the limits of timing and gain of the nTron devices, especially in high-field environments, is becoming a interesting topic to explore. Here we report on results of such investigation. We additionally report on effects of changing the conventional geometry of the nTron by including additional parallel channels to the device current-carrying channel, effectively mimicking the design of superconducting nanowire avalanche detectors~\cite{marsili2011single} in efforts to increase the device speed and gain.

The performance of the nTron device depends primarily on two specifications: the width of the narrow choke at the input gate terminal and the width of the larger bias channel. The ratio between these two dimensions determines the cross-sectional area of the 2D wire and, consequently, the local critical current density $J_c$ \cite{Il'in2005}. The primary function of an nTron is to bias the channel near $J_c$. When a small input signal travels down the gate terminal and into the geometrically constricted choke, it generates a Joule-heated hot spot, increasing the channel's resistivity and resulting in an overall signal amplification. After the input signal switches off, the total current returns below $J_c$, and the hot spot diffuses outward across the channel. The choke area then reverts back to the superconducting state as it is cooled by the surrounding cold bath.

We introduce a modification to the conventional nTron by incorporating nano-patterned gaps, which effectively reduce the width of the biasing channel from the standpoint of switching operation. This transformation converts the wide-channel nTron into a smaller comparable device with lower inductance, while preserving the larger \textit{total} current carrying capacity. The reduced channel inductance leads to improved device recovery time compared to the conventional nTron \cite{Kerman2009, Kerman2006, Toomey2018, Annunziata2010}. Furthermore, this reduction in effective channel width facilitates the fabrication of larger nTron cross sections compared to the traditional choke and bias design, typically ranging from tens to hundreds of nanometers, albeit the larger designs may be limited by the speed of the electro-thermal feedback of the device\cite{McCaughan2014}. 

We explore two different aspect ratios and compare their DC and recovery characteristics in the presence of magnetic fields. We also study effects of patterning the bias channel into parallel nanowires, and provide conventional nTrons of similar aspect ratios for comparison.

Each nTron device was patterned from a 12 nm thick NbN film with a superconducting transition temperature $T_c$ = 7.6 K, deposited using ion beam assisted sputtering on a silicon substrate with a 10 nm silicon nitride underlayer \cite{Polakovic2018}. These nTrons feature a gate input port that narrows into a choke constriction and connects perpendicularly to the primary biasing channel with source and drain ports. The channel near the choke is optimally tapered to maximize current density for improved hot spot sensitivity and reduced current crowding \cite{Clem2011, Inoue2019}. We compare two parallel-channel devices with different aspect ratios to a conventional nTron design.

\begin{figure}
    \centering
    \begin{subfigure}[t]{0.42\textwidth}
        \centering
        \includegraphics[width=\linewidth]{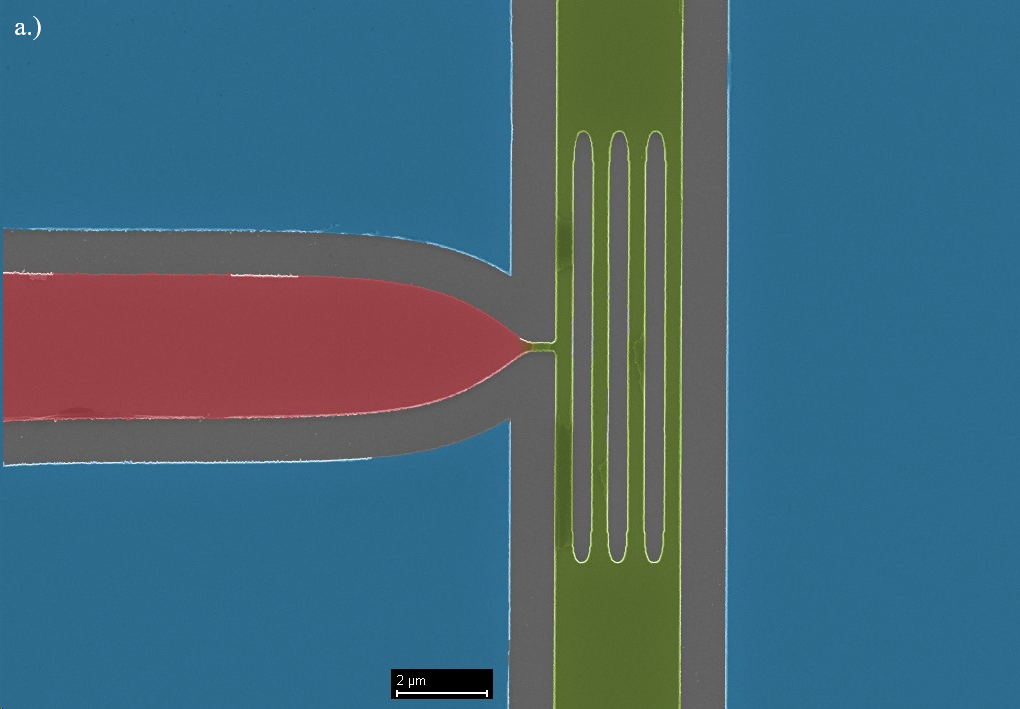} 
    \end{subfigure}
    \begin{subfigure}[t]{0.42\textwidth}
        \centering
        \includegraphics[width=\linewidth]{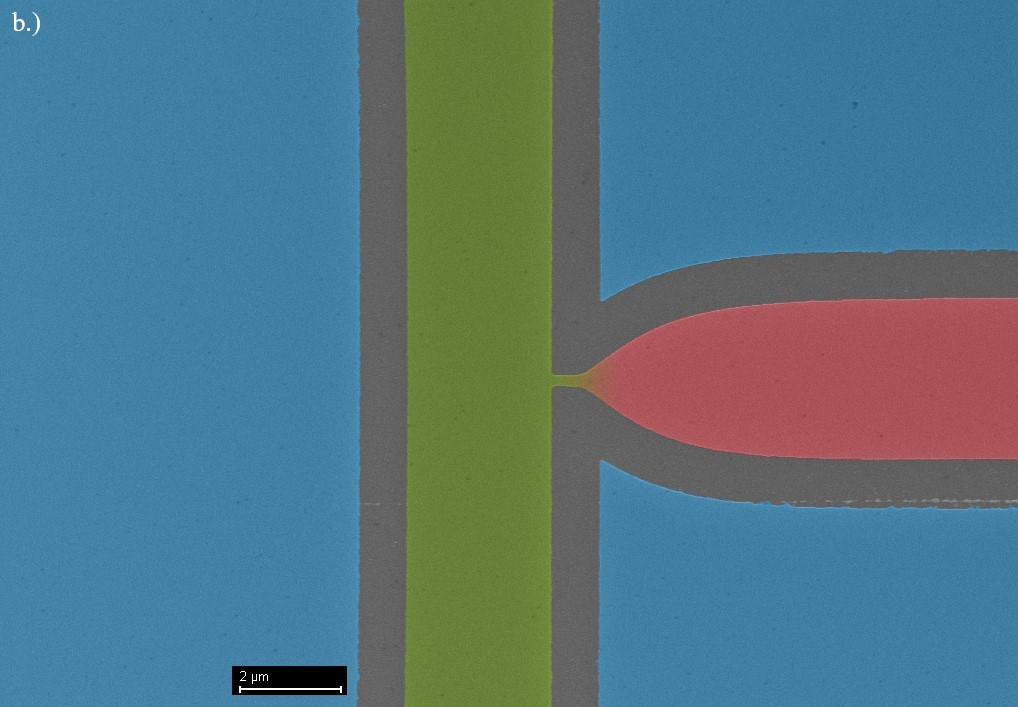} 
    \end{subfigure}

    \caption{False-color SEM images: 1:8 parallel-channel nTron (a), 1:8 conventional nTron (b). Blue highlights the ground plane, grey shows the trench and nanowire gaps, green represents the effective NbN channel, and red signifies the NbN gate to choke constriction. Scale bars correspond to 2 $\mu$m.}
    \label{fig:SEM Both}
\end{figure}

The parallel-channel nTron geometry includes three gaps within the channel of each device, with widths of 100 nm and 400 nm, and lengths 2.4 $\mu$m and 9.6 $\mu$m, respectively. The gaps divide the channel into four parallel wires with smaller cross sectional area near the input gate, effectively reducing the local inductance compared to an nTron with single narrow channel, from $L_{ch}$ to $\frac{L_{ch}}{4}$ because the four narrow channels act as inductors connected in parallel. The choke width is fixed at 200 nm for both designs, while the effective bias channel width is reduced to 400 nm (from 700 nm) and 1.6 $\mu$m (from 2.8 $\mu$m), creating a relative 1:2 and 1:8 aspect ratio between the narrow choke and bias channel widths. In contrast, the conventional nTrons have choke widths of 200 nm and absolute channel widths of 700 nm and 2.8 $\mu$m, respectively. This compares to the size of the nTrons if the gaps were not present within the channel. Fig. \ref{fig:SEM Both} displays scanning electron microscope (SEM) images of both 1:8 parallel-channel and conventional nTron geometries (see supplemental material for 1:2). Table \ref{tab:nTron Specs} summarizes the design specifications for both aspect ratios.

\begin{table}[h]
    \centering
    \begin{tabular}{|c|c|c|c|c|}
    \hline
        nTron & Choke (nm) & Gap (nm) & Eff. Channel ($\mu$m) \\
        \hline
        1:2 (Gapped) & 200 & 100 & 0.4 \\
        \hline
        1:2 (Conv.) & 200 & - & 0.7 \\
        \hline
        1:8 (Gapped) & 200 & 400 & 1.6 \\
        \hline
        1:8 (Conv.) & 200 & - & 2.8 \\
        \hline
    \end{tabular}
    \caption{nTron with gaps feature sizes. Each nTron contains three 100 nm and 400 nm wide gaps placed within the choke area of the channel with respect to each device ratio. The choke spacing shows the width of the gate constriction. The effective channel refers to the net wall-to-wall wire distance of the bias channel.}
    \label{tab:nTron Specs}
\end{table}

We conducted DC characterizations on each nTron device at 1.4 K within an American Magnetics cryostat, which included a superconducting triple-axis vector magnet. We connected the device to an RF chip using aluminum wire bonds and 50 $\Omega$ transmission cables in series with an RF/DC bias tee. To minimize heat dissipation across the channel during switching events, each nTron channel's source and drain ports were connected in parallel to a 50 $\Omega$ shunt \cite{Brenner2012, Toomey2018}.

DC current is initially applied across the channel to characterize the critical switching current of the channel, $I_{c}^{channel}$, indicated by a sharp voltage jump. Once we determined $I_{c}^{channel}$, we bias the channel current within the range of 0.80-1.05 $I_{c}^{channel}$, applying DC current to the nTron gate to identify the critical gate 
 switching current $I_{c}^{gate}$ that initiates the weak link. We then constructed DC switching characteristic diagrams correlating $I_{c}^{gate}$ for weak link generation with the chosen fixed $I^{channel}$. This process was repeated at various magnetic fields $H_{z}$ applied perpendicular to the device surface.

The switching characteristics and output performance of nTron devices depend primarily on the cross-sectional area design of the gate and channel, as well as material properties. In this study, we investigate the impact of magnetic fields on nTron device performance with the parallel-channel modification. By applying DC current to the main nTron channel (without gate input), we observed a sharp voltage rise after the scanning current surpassed the critical switching current $I_{c}^{channel}$. The introduction of a magnetic field produces Abrikosov vortices, shifting the film into a mixed superconducting state. The magnetic field produces a linear increase in residual voltage across the channel resulting from flux flow derived of the Lorentz force interaction of vortices with the biasing current. The field shifts $I_{c}^{channel}$ and reduces the overall amplitude of the voltage jump during the switching transition, since before the jump, there exists some finite voltage provided by the vortex dissipation. As the channel bias approaches $I_{c}^{channel}$, intermittent voltage oscillations cycle out of the mixed superconducting state. This phenomenon is particularly pronounced in the larger 1:8 ratio devices, especially the conventional nTron. Additionally, this effect is accentuated with larger $H_{z}$ due to increased vortex number. Detailed channel and gate scans are provided in the supplemental material.

These intermittent voltage pulses result from vortex crossings in the main nanowire channel of the nTron, either directly or through preexisting hot spots generated by thermal fluctuations in the metastable state induced by higher bias currents \cite{Bulaevskii2011}. Vortex-antivortex pair breaking is unlikely due to the low temperature operation at T = 1.4 K and the higher energy barrier required compared to a single vortex crossing \cite{Kitaygorsky2007}. This phenomenon is especially prominent in larger aspect ratio devices, where the larger cross-section leads to a higher $I_{c}^{channel}$ threshold and, consequently, higher operation bias, even without an applied magnetic field \cite{McCaughan2014}.

\begin{figure}
    \centering
    \begin{subfigure}[t]{0.45\textwidth}
        \centering
        \includegraphics[width=\linewidth]{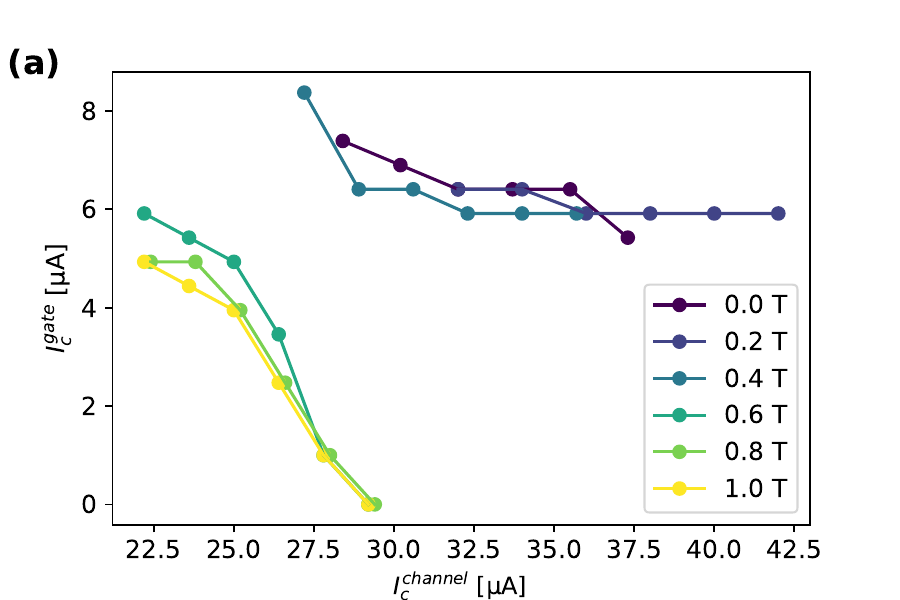} 
    \end{subfigure}
    \hfill
    \begin{subfigure}[t]{0.45\textwidth}
        \centering
        \includegraphics[width=\linewidth]{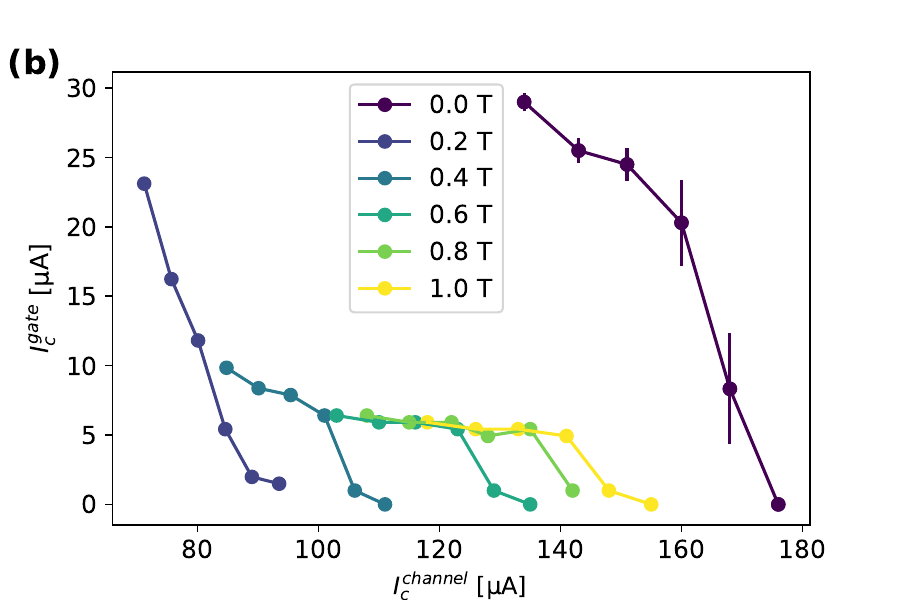} 
    \end{subfigure}

    \caption{1:2 (a) and 1:8 (b) parallel-channel nTron DC switching characteristic diagrams in magnetic fields. The 1:8 device at $H_{z} = 0$ demonstrates stochastic switching behavior and is reported as averages with error bars denoting one standard deviation.}
    \label{fig:nTron Islands 14 and 116 PD}
\end{figure}

Fig.~\ref{fig:nTron Islands 14 and 116 PD} shows the critical current switching relationship between the channel bias and input gate signal for both the 1:2 and 1:8 parallel-channel nTrons. Notably, the 1:2 parallel-channel nTron exhibits distinct behavior at different magnetic field strengths, characterized as "low fields" (0-0.4 T) and "high fields" (0.6-1.0 T), respectively. This suggests two regimes: one dominated by current and the other by vortices. In the low-field regime, where vortex population is smaller and flux motion weaker, the weak link transition is less sensitive to a stronger input pulse. However, as the magnetic field strength increases, the vortex interaction becomes strongly coupled to the biasing current, leading to a downshift in the critical current and increased sensitivity to the gate input signal. 

\begin{figure}[h]
    \centering
    \begin{subfigure}[t]{0.45\textwidth}
        \centering
        \includegraphics[width=\linewidth]{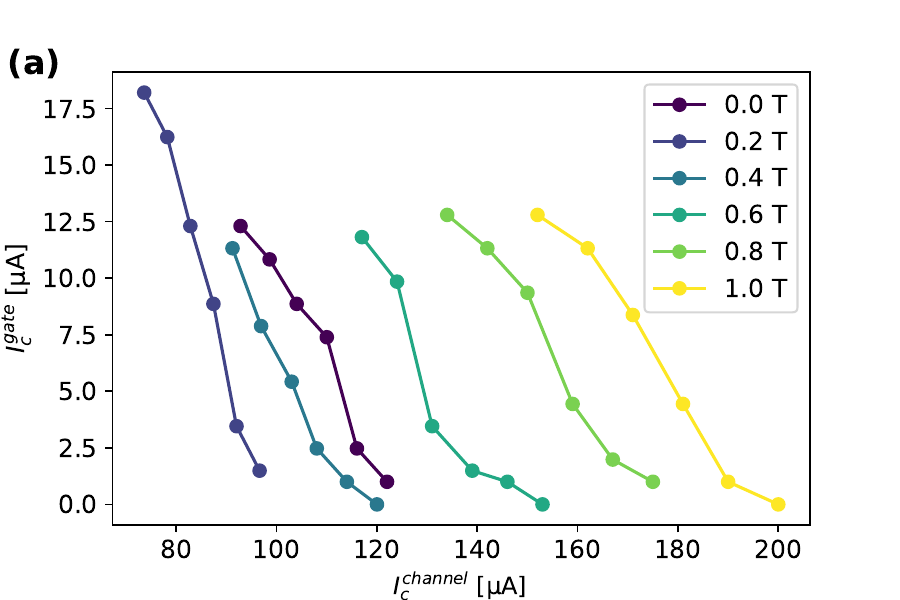} 
        \label{fig:1:4 nTron no Islands PD}
    \end{subfigure}
    \hfill
    \begin{subfigure}[t]{0.45\textwidth}
        \centering
        \includegraphics[width=\linewidth]{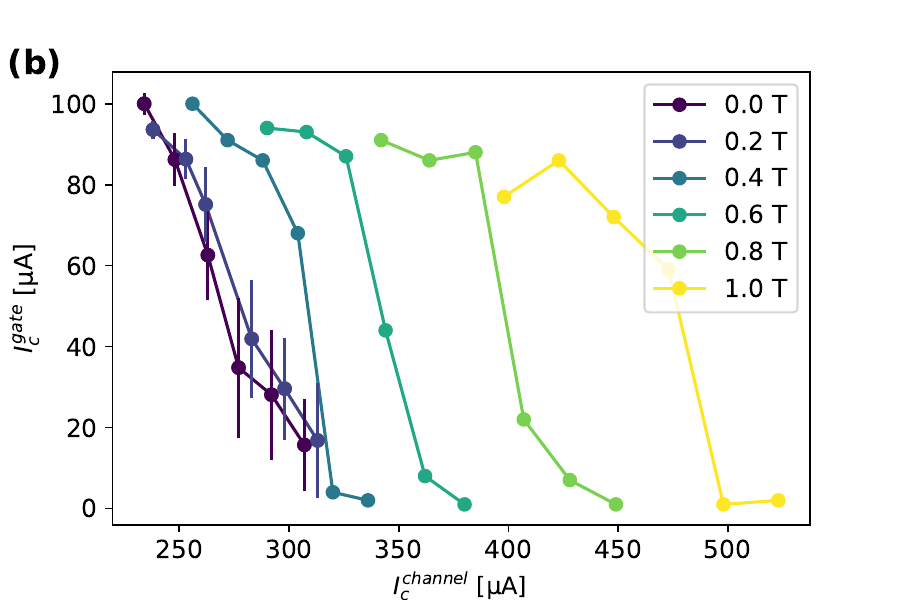}
    \end{subfigure}

    \caption{1:2 (a) and 1:8 (b) conventional nTron DC switching characteristic diagrams in magnetic fields. The 1:8 device at $H_{z} = 0$ and $H_{z} = 0.2$ T demonstrates stochastic switching behavior and is reported as averages with error bars denoting one standard deviation.}
    \label{nTron no Islands 14 and 116 PD}
\end{figure}

As the nTron's aspect ratio increases, the regime shift becomes less pronounced. In Fig.~\ref{fig:nTron Islands 14 and 116 PD}, the 1:8 nTron displays a distinct shift in gate signal sensitivity when a magnetic field is applied, causing $I_{c}^{channel}$ to decrease and gate sensitivity to plateau around 5 µA. Due to its larger primary channel, the 1:8 device exhibits stochastic switching events within a window of $I_{c}^{channel}$ values. This behavior is also observed in the 1:8 conventional device, even with a small 0.2 T applied field (see Fig. \ref{nTron no Islands 14 and 116 PD}). The more prominent switching observed in the conventional nTron is attributed to its larger channel cross-sectional area. The nano-patterned gaps, however, lower the local maximum  vortex count, foster pinning, and lessen the probabilistic $I_{c}^{channel}$ switching, as evident in the 1:8 design \cite{Colauto2021, Wang2013}.

It is noteworthy to observe the increase in 
$I_{c}^{channel}$ for the larger 1:2 conventional device with a wider channel compared to its parallel-channel twin device. This observation is also consistent for both 1:8 nTron designs. Previously, we described how the application of $H_{z}$ leads to a linear increase in voltage with scanning current across the channel. With increasing $H_{z}$, the voltage jump, signaling the nTron transitioning from a mixed state to fully normal state, diminishes in amplitude and appears at higher $I^{channel}$ (see Fig. S1-2 supplemental material). We focus at this distinct voltage jump which indicates a complete transition as the hotspot fully encloses the channel, opposed to the low-bias regimes where the residual voltage signal gradually increases from the zero level.

From these observations, it is inferred that the escalating field strength intensifies flux flow jamming, leading to an increased channel switching current \cite{Karapetrov2012}. This escalation results in a residual voltage increase, attributable to flux flow induced by the biasing current. As the field strength increases, higher vortex density enhances the depinning force, which in turn affects the vortex mobility, leading to jamming and altered dynamics. This relationship lays the foundation for the observed voltage behavior. Initially, the voltage remains at zero. With the application of a field and the resultant screen current, a linear increase in voltage is witnessed due to flux flow. However, the sharp voltage jump marking $I_{c}^{channel}$ shifts to higher values with each increase in field strength, and the amplitude of this "jump" concurrently diminishes. This behavior is ascribed to the intricate interplay between the driving current and the distorted vortex lattice. At heightened field strengths, the entire vortex lattice, now denser and more constricted, moves under the influence of the driving bias, leading to a reduced voltage gain at $I_{c}^{channel}$ and altered vortex dynamics \cite{Reichhardt2010}.

After determining the DC characteristics of the nTron, we measured and compared its recovery speed using two separate gate pulse techniques. Nanoseconds wide pulses were delivered to the gate via a pulse generator at variable attenuation, as the input pulse power controls the hotspot size and temperature \cite{Zheng2019}. The first technique involved a double short pulse train, consisting of two brief pulses. The timing of these pulses was adjusted so that the second pulse was delivered at approximately 50\% decay of the voltage signal from the first pulse. The second technique measured the step response from a long square pulse. The pulse length was 50 ns to minimize reflections and other transient effects. We measured the voltage signal across the channel using a Lecroy oscilloscope connected by a bias tee. The time constant $\tau$ was extracted from the decaying tail of the voltage signal using an exponential decay fitting function: $V(t) = V_{0}e^{-\frac{t}{\tau}}$. This procedure was repeated at different amplitudes of $H_{z}$ up to 1.0 T. 

Pulse and attenuation parameters were scanned to optimize the signal-to-noise ratio while minimizing any latching caused by reflections or over stimulation of the gate \cite{Zheng2019}. The primary pulse parameters included applied gate voltage, pulse attenuation, pulse duration, and pulse spacing (for the double pulse only). Detailed pulse parameters for each nTron device in its respective field environment are provided in the supplemental material. The maximum voltage output varied in different bias and field environments from the trade off of input power to recovery time. In addition, some latching and reflection was unavoidable at fixed attenuation in order to have distinguishable signal-to-noise ratio, making the measurement of the actual recovery difficult in some cases. 

\begin{figure}[h]
    \centering
    \begin{subfigure}[t]{0.45\textwidth}
        \centering
        \includegraphics[width=\linewidth]{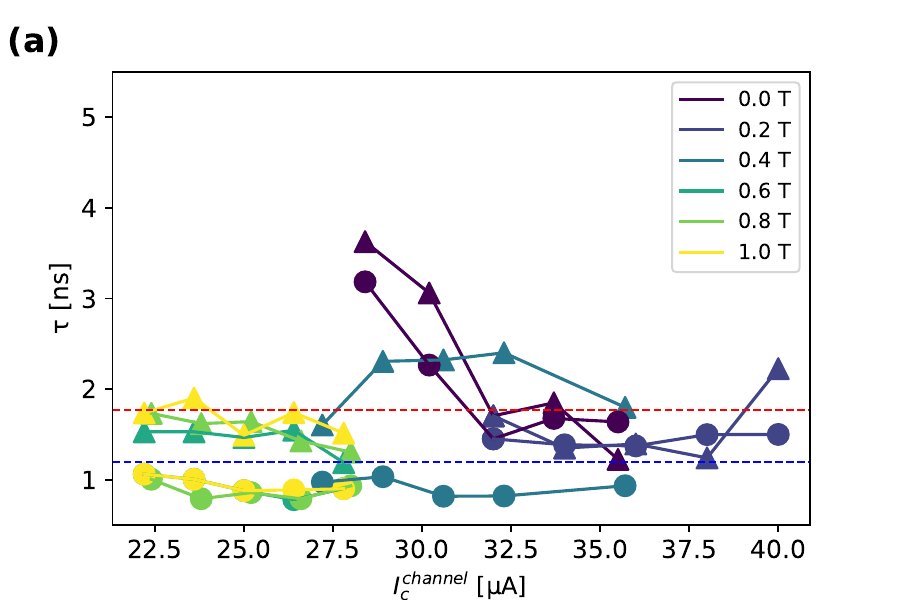} 
    \end{subfigure}
    \hfill
    \begin{subfigure}[t]{0.45\textwidth}
        \centering
        \includegraphics[width=\linewidth]{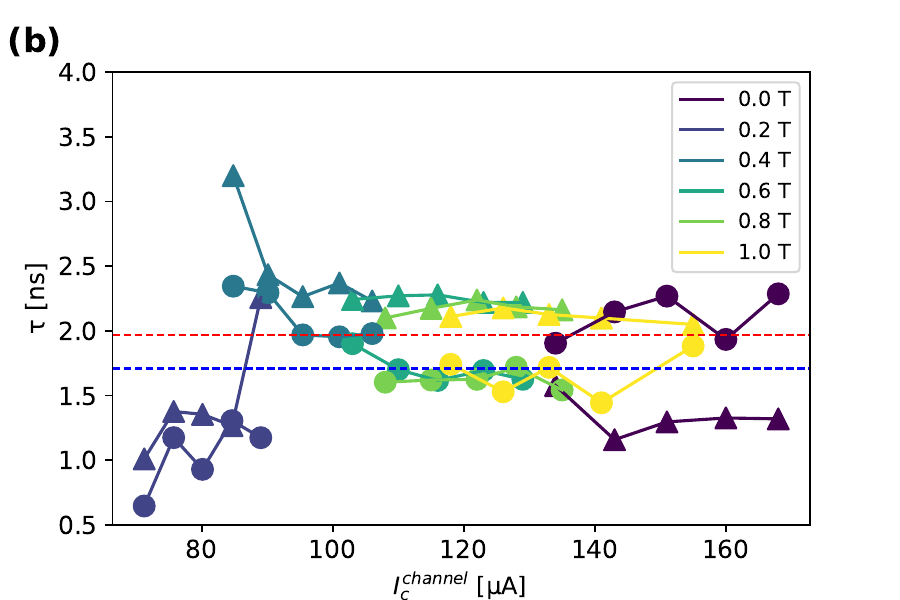} 
    \end{subfigure}

    \caption{1:2 (a) and 1:8 (b) parallel-channel nTron thermal relaxation times for both double (triangles) and step response (circles) pulse methods. Dashed lines indicate the average for each technique: double pulse (red) and step response (blue).}
    \label{fig:nTron Islands Relax Time}
\end{figure}

Fig.~\ref{fig:nTron Islands Relax Time} shows the relaxation times for biases at 0.80-1.0 $I_{c}^{channel}$ of the 1:2 and 1:8 nTrons with parallel channels. The longer relaxation times at low bias and zero field are likely due to direct cooling from the substrate, rather than the faster dynamics of the superconductor in high-field environments \cite{Kerman2009}.

\begin{figure}[h]
    \centering
    \begin{subfigure}[t]{0.45\textwidth}
        \centering
        \includegraphics[width=\linewidth]{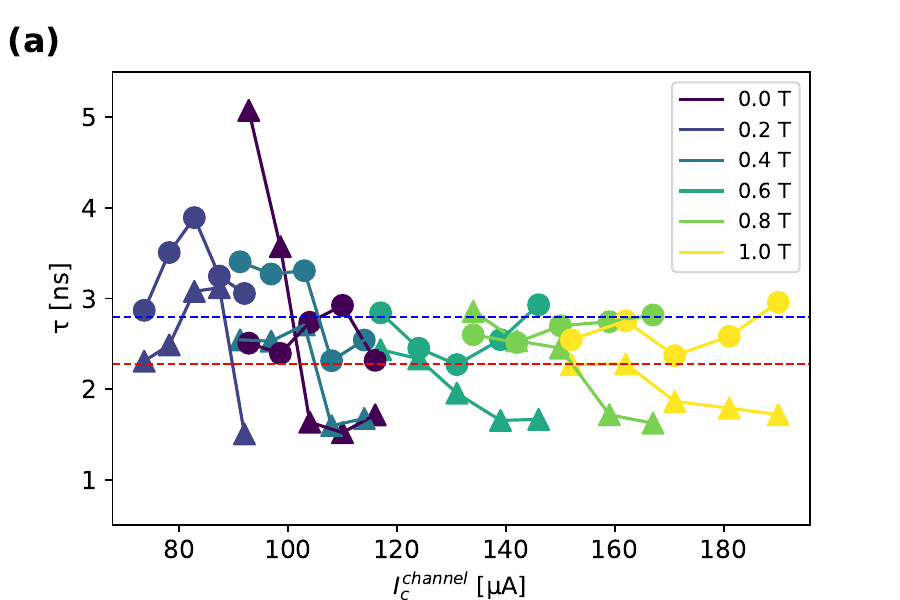} 
    \end{subfigure}
    \hfill
    \begin{subfigure}[t]{0.45\textwidth}
        \centering
        \includegraphics[width=\linewidth]{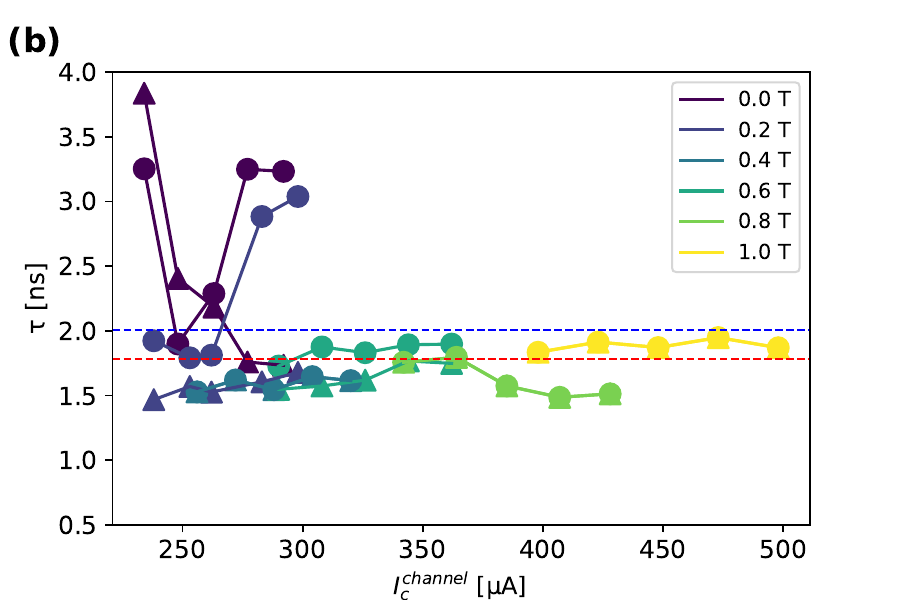} 
    \end{subfigure}

    \caption{1:2 (a) and 1:8 (b) conventional nTron thermal relaxation times for both double (triangles) and step response (circles) pulse methods.Dashed lines indicate the average for each technique: double pulse (red) and step response (blue).}
    \label{fig:nTron No Islands Relax Time}
\end{figure}

Conversely, Fig.~\ref{fig:nTron No Islands Relax Time} shows the recovery speeds of the conventional nTrons. The 1:2 device demonstrates a slower speed on average when compared to its gapped counterpart by 0.5 ns (double) and 1.61 ns (step) between the two pulse methods. Whilst, the larger 1:8 scale design only shows an average increase of 0.29 ns (step) and notably a decrease of -0.19 ns (double) in $\tau$ between the two methods. There is large variation at $H_{z} = $ 0 and 0.2 T derived from the stochastic nature of the switching seen in the DC characteristics, but is otherwise consistent between both pulsing techniques. Table ~\ref{tab:Tau Averages} indicates the averages of each technique for all devices including the difference $\Delta$ between the gapped and conventional design.

\begin{table}[ht]
    \centering
    \begin{tabular}{|c|c|c|c|c|c|c|}
    \hline
    \textbf{Ratio} & \multicolumn{3}{c|}{\textbf{1:2 (ns)}} & \multicolumn{3}{c|}{\textbf{1:8 (ns)}} \\
    \cline{2-7}
    \hline
      & \textbf{Gapped} & \textbf{Conv.} & \textbf{$\Delta$} & \textbf{Gapped} & \textbf{Conv.} & \textbf{$\Delta$} \\
    \hline
    Double & 1.77 & 2.27 & 0.50 & 1.97 & 1.78 & 0.19 \\
    \hline
    Step & 1.19 & 2.80 & 1.61 & 1.71 & 2.00 & 0.29 \\
    \hline
    \end{tabular}
    \caption{Thermal relaxation averages (in ns) for each aspect ratio and respective pulse method. $\Delta$ indicates the difference in averages between the gapped and conventional nTron designs.}
    \label{tab:Tau Averages}
\end{table}

In summary, our study demonstrates larger-size NbN nTron operation through the introduction of parallel connection of multiple channels, increasing the effective channel width. These modifications reduce stochastic switching events, enhance gate signal sensitivity, and promote vortex pinning which helps reduce false signal noise, particularly at higher bias currents. While reducing the channel width has a minimal impact on recovery times of the 1:8 devices with wider aspect ratios, the smaller 1:2 ratio design exhibits enhanced recovery speeds on average as fast as 1.19 ns compared to its conventional counterpart of 2.80 ns, making it suitable for on-chip integration with other superconducting circuitry like SNSPDs. Future work may involve adjusting channel gap sizes and length in 1:2 ratio designs to further optimize gate sensitivity at higher bias currents and improve recovery times at high current gains.

See the supplementary material for SEM images of nTrons (Fig. S1), DC channel and gate switching current characterization scans (Fig. S2-7 and Table S1-3), and recovery time pulse parameters (Table S4-5).

\begin{acknowledgments}
This work was supported by the U. S. Department of Energy (DOE), Office of Science, Office of Nuclear Physics, Microelectronics Initiative, under Contract No. DE-AC02-06CH11357. Y. L. acknowledges support by the U.S. DOE, Office of Science, Office of Basic Energy Sciences, Materials Sciences and Engineering Division under Contract No. DE-SC0022060. Work performed at the Center for Nanoscale Materials, a U.S. Department of Energy Office of Science User Facility, was supported by the U.S. DOE, Office of Basic Energy Sciences, under Contract No. DE-AC02-06CH11357.
\end{acknowledgments}

\section*{Author Declarations}
\section*{Conflict of Interest}

The authors have no conflicts to disclose.

\section*{Author Contributions}

\textbf{Timothy Draher:} Conceptualization (equal); Data curation (lead); Formal analysis (lead); Investigation (lead); Methodology (equal); Resources (supporting); Software (lead); Validation (equal); Visualization (lead); Writing -  original draft (lead); Writing - review and editing (lead). \textbf{Tomas Polakovic:} Conceptualization (lead); Formal analysis (equal); Methodology (lead); Project Administration (equal); Resources (supporting); Software (supporting); Supervision (equal); Validation (lead); Visualization (supporting); Writing -  original draft (supporting); Writing - review and editing (supproting);  \textbf{Yi Li:} Methodology (supporting); Supervision (supporting); Validation (supporting). \textbf{John Pearson:} Supervision (supporting). \textbf{Alan Dibos:} Methodology (supporting); Resources (supporting). \textbf{Zein-Eddine Meziani:} Project Administration (supporting); Supervision (supporting). \textbf{Zhili Xiao:} Supervision (supporting). \textbf{Valentine Novosad:} Funding acquisition (lead); Methodology (supporting); Project administration (lead); Resources (lead); Supervision (equal); Validation (equal);

\section*{Data Availability}

The data that support the findings of this study are available within the article and its supplementary material.

\nocite{*}
\section*{References}
\bibliography{aipsamp}% Produces the bibliography via BibTeX.

\end{document}